\documentclass[%
 aps,prl,showpacs,
 amsmath,amssymb,
 reprint,%
]{revtex4-1}

\usepackage{graphicx}
\usepackage{dcolumn}
\usepackage{bm}
\usepackage{color}

\graphicspath {
    {./}
    {./Figures/}
}

\begin{document}

\preprint{AIP/123-QED}

\title{Piezoelectric mimicry of flexoelectricity}

\author{Amir Abdollahi}
\affiliation{Laboratori de C\`alcul Num\`eric (LaC\`aN), Universitat Polit\`ecnica de Catalunya (UPC), Campus Nord UPC-C2, E-08034 Barcelona, Spain.}%

\author{Fabi\'an V\'asquez-Sancho }

\affiliation{Catalan Institute of Nanoscience and Nanotechnology (ICN2), CSIC and The Barcelona Institute of Science and Technology, Campus UAB, Bellaterra, 08193 Barcelona, Spain}%
\affiliation{Centro de Investigaci{\'o}n en Ciencia e Ingeniería de Materiales, Universidad de Costa Rica, San Jos{\'e} 11501, Costa Rica}

\author{Gustau Catalan}
\email{gustau.catalan@icn2.cat}
\affiliation{Catalan Institute of Nanoscience and Nanotechnology (ICN2), CSIC and The Barcelona Institute of Science and Technology, Campus UAB, Bellaterra, 08193 Barcelona, Catalonia}%
\affiliation{ICREA-Institut Catala de Recerca I Estudis Avan\c cats, Barcelona, Catalonia}%

\date{\today}

\begin{abstract}
The origin of ``giant'' flexoelectricity, orders of magnitude larger than theoretically predicted, yet frequently observed,
is under intense scrutiny. There is mounting evidence correlating giant flexoelectric-like
effects with parasitic piezoelectricity, but it is not clear how piezoelectricity (polarization generated by strain) 
manages to imitate flexoelectricity (polarization generated by strain gradient) in typical beam-bending experiments, since in a bent beam the net strain
is zero. In addition, and contrary to flexoelectricity, piezoelectricity changes sign under space inversion, 
and this criterion should be able to distinguish the two effects and yet ``giant'' flexoelectricity is insensitive to space inversion, seemingly contradicting a piezoelectric origin. Here we show that, if a piezoelectric material has
its piezoelectric coefficient be asymmetrically distributed across the sample, it will generate a bending-induced 
polarization impossible to distinguish from “true” flexoelectricity even by inverting the sample. The effective
flexoelectric coefficient caused by piezoelectricity is functionally identical to, and often larger than, intrinsic
flexoelectricity: the calculations show that, for standard perovskite ferroelectrics, even a tiny gradient of piezoelectricity
(1\% variation of piezoelectric coefficient across 1 mm) 
is sufficient to yield a giant effective flexoelectric coefficient
of 1 $\mu$C/m, three orders of magnitude larger than the intrinsic expectation value.

\end{abstract}


\maketitle

Flexoelectricity is attracting growing attention due to its ability to replicate the electromechanical functionality of piezoelectric materials, 
which opens up the possibility of using lead-free dielectrics as flexoelectric replacements for piezoelectrics in specific applications\cite{ref:Chu2009,ref:Bhaskar2015}. 
Experimental research on this phenomenon is still in a relative infancy, but already there have been controversies about the real magnitude, origin and even thermodynamic reversibility of the flexoelectric effect \cite{ref:Yudin2013,ref:Zubko2013,ref:Nguyen2013}. 
Some of these controversies are starting to get settled, and, in particular, there is by now abundant evidence and growing consensus that seemingly ``giant'' flexoelectric effects are correlated with parasitic piezoelectric contributions from polar nanoregions \cite{ref:Navarez2014}, defect concentration gradients \cite{ref:Biancoli2015}, residual ferroelectricity \cite{ref:Garten2015}, or surfaces \cite{ref:Tagantsev1986, ref:Narvaez2015, Narvaez2016219}. But, while the recent evidence suggests that indeed piezoelectricity
can mimic flexoelectricity (which is the converse of flexoelectricity replicating piezoelectricity), it is not clear how (i.e., what are the necessary conditions for piezoelectricity to be able to imitate flexoelectricity), nor to what extent is the ``disguise'' perfect, i.e., can intrinsic flexoelectricity and flexoelectric-like piezoelectricity be experimentally distinguished? 

\begin{figure*}[!htp]
\centering
\includegraphics[clip, angle=0, width=1\textwidth]{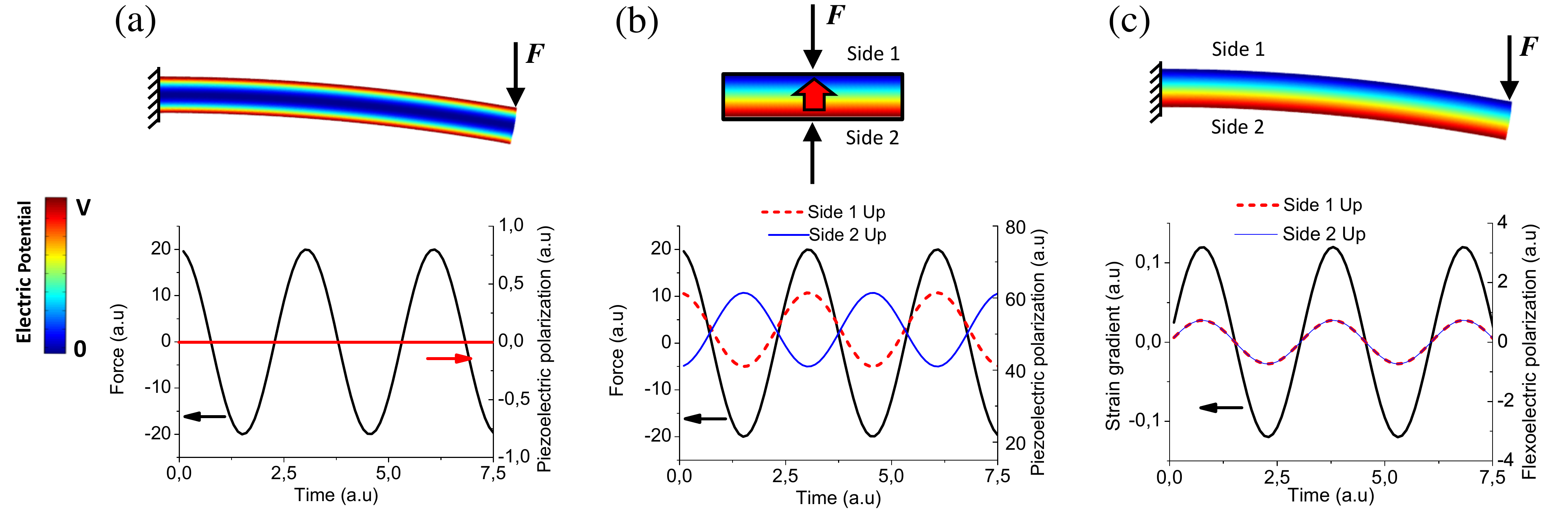}
\caption{ (a) Homogeneously poled piezoelectric beam under bending, which does not induce a non-zero net polarization because the average strain is zero. 
The color plot presents the electric potential distribution. (b) Piezoelectric polarization induced in a rectangular sample under tension or compression. 
The red arrow represents the direction of the material polarization. (c) Flexoelectric polarization induced in a cantilever beam under bending. 
The polarization does not change sign by reversing the beam.}\label{fig1}
\end{figure*}

To illustrate these questions, consider the following example: polarization can be generated by flexoelectricity when the applied deformation is inhomogeneous, e.g., when a sample is bent \cite{ref:Bursian1968,ref:Ma2005,ref:Cross2006,ref:Zubko2007,ref:Majdoub2008B,ref:Kwon2014}, but this is not necessarily true for piezoelectricity: bending a homogeneously poled piezoelectric beam will not elicit any piezoelectric polarization, see Fig.~\ref{fig1} (a), because there is no net strain: the piezoelectric polarization caused by stretching on the convex side will be canceled by
the opposite polarization caused by compression on the concave side. It follows from this example that the apparently giant flexoelectricity measured in the bending of some materials 
cannot be caused by a homogeneous piezoelectric state, even if the sample is macroscopically polar \cite{ref:Biancoli2015}. Furthermore, because the existence of macroscopic piezoelectricity can be established by space-inversion experiments such as flipping the sample upside-down and verifying that the sign of the stress-generated charge changes sign \cite{ref:Biancoli2015}, see Fig.~\ref{fig1} (b), it was assumed that such space-inversion tests could be used to distinguish between piezoelectricity and flexoelectricity \cite{ref:Narvaez2015}. Indeed, the bending-induced polarization of a flexoelectric cantilever is independent of its orientation, see Fig.~\ref{fig1}(c), but, as we will see, this inversion invariance can also hold for bent piezoelectric cantilevers.  

Here, we analyze the electromechanical response of a bent piezoelectric beam, one of the common setups to quantify flexoelectricity \cite{ref:Zubko2013},  
and show that (i) it is a necessary and sufficient condition that the piezoelectric coefficient be asymmetrically distributed for the beam to be able to replicate 
the functional behavior of a flexoelectric and (ii) that such asymmetric piezoelectricity cannot be distinguished from flexoelectricity in beam-bending experiments, even if the sample is turned upside down; the disguise is, in this respect, perfect. 
It is also possible to define an effective flexoelectric constant as a function of the spatial distribution of piezoelectricity. Quantitative analysis of this piezo-flexoelectric coefficient shows that even a relatively modest asymmetry in the distribution of piezoelectricity can lead to an effectively giant flexoelectric effect.

\begin{figure*}
\centering
\includegraphics[clip, angle=0, width=0.8\textwidth]{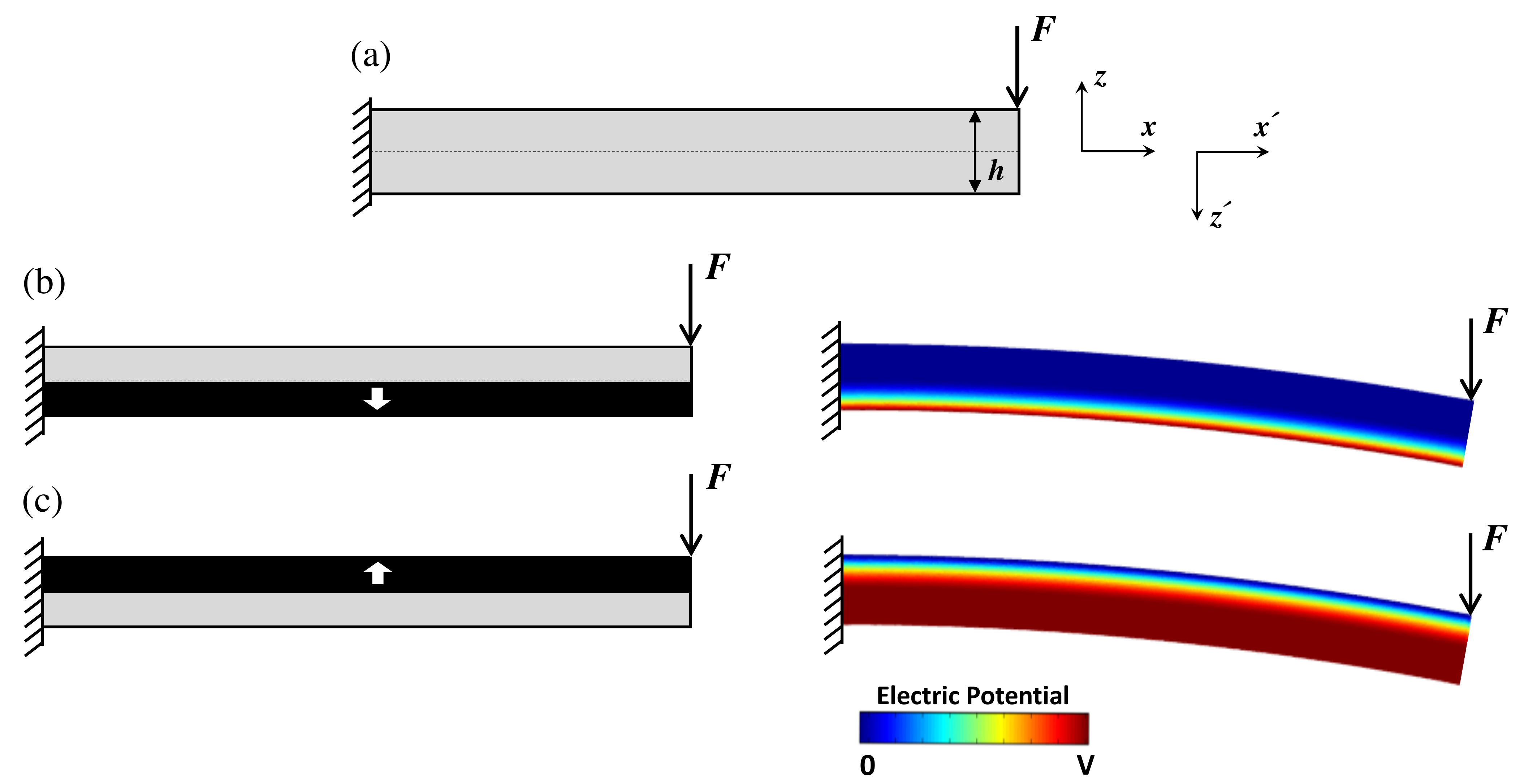}
\caption{  (a) Schematic of a cantilever beam under the point load $F$. (b) Schematic of a bimorph cantilever beam where the dark layer is piezoelectric. 
(c) A reversed configuration of the bimorph (rotated by 180$^o$). The white arrows represent the polarization direction of the layers. The color plot 
presents the electric potential ($\phi$) distribution for each bimorph obtained from Eqs.~\eqref{equiD0} and \eqref{equiPz}, where $E_z = -\phi_{,z}$. The net polarization can be obtained from the electric potential difference
of the top and bottom faces, which is identical in both bimorphs. }\label{fig3}
\end{figure*}

The constitutive equation for the electric displacement $\mathbf{D}$ in a linear dielectric solid possessing piezoelectricity and flexoelectricity is
\begin{equation}
D_i =  e_{ikl} \varepsilon_{kl} + \mu_{ijkl} \nabla_l \varepsilon_{jk} + \epsilon_{ij}E_j =  \epsilon_{0}E_i + P_i , \label{equiD}
\end{equation}
where $\mathbf{E}$ is the electric field, $\mathbf{\varepsilon}$ is the mechanical strain, $\mathbf{P}$ is the polarization, $\nabla \mathbf{\varepsilon}$ is the strain gradient,
 $\mathbf{e}$ is the piezoelectric tensor, $\mathbf{\mu}$  is the flexoelectric tensor, $\mathbf{\epsilon}$ is the dielectric tensor, and $\epsilon_0$ is the permittivity of vacuum or air. We begin by analyzing the response of a
piezoelectric flexoelectric cantilever beam under bending, see Fig.~\ref{fig3}(a). We assume that the electric field and polarization exist only in the 
beam thickness direction ($z$) since it has been shown that the longitudinal electric filed is negligible compared with that in the beam thickness direction \cite{Gao2007, Wang2010}. Then, Eq.~\eqref{equiD} simplifies to
\begin{equation}
D_z =  e_{31} \varepsilon_{xx} + \mu_{13} \varepsilon_{xx,z} + \epsilon_{33}E_z =  \epsilon_{0}E_z + P_z , \label{equiDS}
\end{equation}
where the notations $e_{311} = e_{31}$ and  $\mu_{1331} = \mu_{13}$ are introduced for convenience.

In the absence of surface charges and applied voltage, the electrostatic equilibrium (Maxwell's equation) leads to 
\begin{equation}
D_z  =  \epsilon_{0}E_z + P_z = 0. \label{equiD0}
\end{equation}

Plugging this equation in Eq.~\eqref{equiDS} and using the Euler beam hypotheses $\varepsilon_{xx} = -\kappa z$ and $\varepsilon_{xx,z} = -\kappa$, 
where $\kappa$ is the beam curvature induced by the applied force $F$, the polarization in $z$ direction can be obtained as
\begin{equation}
P_z (z) = - \frac{\kappa}{\epsilon_r} (e_{31} z + \mu_{13}), \label{equiPz}
\end{equation}
where $\epsilon_r = \epsilon_{33}/\epsilon_0$ is the relative dielectric constant. We note that this equation can also be derived from analytical solutions of the electroelastic fields 
in bending piezoelectric cantilever beams with the flexoelectric effect \cite{ref:Majdoub2008B, ref:Majdoub2009, ref:Yan2013-a} or gradient piezoelectricity \cite{ref:Williams1982, ref:Williams1983}. The total net polarization over the beam thickness is then obtained as

\begin{equation}
P_t =   \frac{1}{h}\int_{-h/2}^{h/2} P_z dz = - \frac{\kappa}{h\epsilon_r}\int_{-h/2}^{h/2} (e_{31} z + \mu_{13}) dz. \label{equiPt}
\end{equation}

In the absence of piezoelectricity, i.e. $e_{31} = 0$, the polarization is only induced by the flexoelectric effect, resulting in a net polarization $P_t = - \mu_{13} \kappa / \epsilon_r $, 
independent of the beam direction. In other words, the net polarization induced by flexoelectricity does not change sign by reversing the beam, as expected.

We can also use Eq.~\eqref{equiPt} to obtain the net polarization corresponding to a beam where the electromechanical response is piezoelectric instead of flexoelectric, i.e. $\mu_{13} = 0$. However,
bending a homogeneous piezoelectric beam is unable to produce 
a non-zero net polarization since $e_{31}z$ is antisymmetric about the centre of the beam ($z$ = 0), 
thus leading to a zero net polarization in Eq.~\eqref{equiPt}, see Fig.~\ref{fig1}(a). The physical reason, as mentioned at the introduction, is that opposite stresses (compressive and tensile) are induced in the upper
 and lower halves of the beam respectively, resulting in opposite piezoelectric effects and a zero net polarization. Therefore, 
 bending a homogeneous piezoelectric beam cannot give a flexoelectric-like response.

One way to break the balance of charges in a bent piezoelectric beam is to replace the top 
or bottom layers with a different piezoelectric or even a non-piezoelectric layer, as is done for example in piezoelectric bimorph sensors and actuators \cite{ref:Smits1991,ref:Pritchard2001}. 
One such bimorph is illustrated in Fig.~\ref{fig3}(b), in which the bottom layer is piezoelectric while the top layer is not, i.e. $e_{31} = 0$ for $h\geq0$ and $e_{31} = -e_{31} $ for $h<0$. 
The bimorph can be seen as an extreme case of asymmetric piezoelectricity, where $e_{31}$ is a Heaviside step function. In this case the net polarization is obtained from Eq. \eqref{equiPt} as $P_t$ = $- e_{31} h \kappa /8 \epsilon_r $.

A bimorph piezoelectric cantilever thus generates a polarization just like a flexoelectric cantilever would. Moreover, the sign (phase shift) of the piezoelectric polarization 
does not change by reversing the beam. Figure \ref{fig3}(c) presents the reversed configuration which is equivalent to consider $e_{31} = 0$ for $h<0$. 
 Plugging these conditions in Eq.~\eqref{equiPt} leads to a net polarization  
$P_t$ = $- e_{31} h \kappa /8 \epsilon_r $, identical both in magnitude and sign, to the induced polarization in the original bimorph in Fig.~\ref{fig3}(b). 
Therefore, a bent piezoelectric bimorph is qualitatively indistinguishable from a bent flexoelectric beam.

We can generalize the conclusions of this example. Let us assume a generic cantilever beam with an arbitrary distribution of piezoelectricity $e_{31}(z)$.  Equation \eqref{equiPt} results in a zero net polarization
if the piezoelectricity is symmetrically distributed about the centre of the beam, i.e. for any $e_{31}(z)$ such that $e_{31}(z) = e_{31}(-z)$. Mathematically
the integrand is antisymmetric about the centre of the beam for any such symmetrical distribution of piezoelectricity. 
Conversely, any asymmetry in the distribution of piezoelectricity such that $e_{31}(z) \ne e_{31}(-z)$ will result in a non-zero integral and thus in a net bending-induced piezoelectric polarization.

In addition, the sign of the net polarization in Eq.~\eqref{equiPt}
does not change by flipping the beam. In the flipped configuration, the coordinate system $x - z$ converts to the new system $x^\prime - z^\prime$, where $z^\prime = - z$.  Using this conversion and taking into account the negative sign of $e_{31}$ in the flipped configuration, 
Eq.~\eqref{equiPt} converts to an identical equation as a function of $z^\prime$, retaining its sign. Therefore,
for a piezoelectric beam to be able to indistinguishably mimic a flexoelectricity (i.e., for Eq.~\eqref{equiPt} to yield a non-zero solution that 
is invariant with respect to space inversion), it is necessary and sufficient that the piezoelectric coefficient be asymmetrically distributed across the thickness of the beam. Two particular embodiments of this general concept are the bimorph piezoelectric cantilever, for which $e_{31}(z)$ is a step-function, and surface piezoelectricity \cite{ref:Tagantsev2012,ref:Zubko2013,ref:Stengel2014,ref:Narvaez2015}, for which $e_{31}(z)$ can be viewed as two step functions. 

Since an asymmetric piezoelectric can identically mimic a flexoelectric-like response, it is possible to define an effective flexoelectric constant as a function of the 
distribution of piezoelectricity. For a flexoelectric cantilever, the induced polarization as a function of the beam curvature is given by $P_t = - \mu_{e} \kappa / \epsilon_r $, where $\mu_{e}$ is the effective flexoelectric constant. By equating this polarization to Eq. \eqref{equiPt} with $\mu_{13} = 0$, the effective flexoelectric constant becomes 

\begin{equation}
\mu_e \equiv  \frac{1}{h} \int_{-h/2}^{h/2} {e}_{31}(z) z dz. \label{equi5}
\end{equation}

In order to get some quantitative estimates of how much “pseudo-flexoelectricity” can we elicit from a gradient of piezoelectricity, we consider a simple linear distribution of piezoelectricity as
$e_{31}(z) = z {\Delta e} /h + e_0$, where $\Delta e/h$ is the slope of the linear gradient of piezoelectricity and $e_0=e_{31}(0)$. Plugging this function 
in Eq. \eqref{equi5} yields an effective flexoelectric coefficient of

\begin{equation}
\mu_e = h \Delta e/12. \label{equi6}
\end{equation}

Let us use this equation to analyze relevant experimental cases. Experimental setups to quantify flexoelectricity commonly employ cantilever beams
with a thickness in the order of $h = 1$ mm \cite{ref:Ma2001,ref:Ma2002,ref:Ma2005,ref:Ma2006,ref:Cross2006,ref:Shu2013}, so their piezoelectrically-induced flexoelectricity would be $\mu_e \approx \Delta e\times10^{-4}$. Therefore, to induce a typical ``giant'' flexoelectric coefficient in the order of
$\mu_e =$ 1 $\mu$ C/m, as reported for important piezoelectric materials such as PZT and BaTiO$_3$ \cite{ref:Ma2005,ref:Cross2006,ref:Ma2006}, the piezoelectric variation $\Delta e$ between the two sides of the 1 mm sample 
should be in the order of $10^{-2}$ C/m$^2$. Compared to the average piezoelectric coefficient of PZT and BaTiO$_3$, which
is in the order of 5 C/m$^2$ \cite{ref:Li1991,ref:Zhu1998b}, this gradient is equivalent to a 0.2 \% change of the piezoelectric constant across the beam thickness. 
Therefore, for materials with big piezoelectric coefficients, 
even a tiny gradient of piezoelectricity can yield an apparently giant flexoelectricity. 
This invalidates bending-based quantifications of flexoelectricity in the polar phase of these materials and highlights 
the need for alternative experimental approaches \cite{Cordero2017}.

Another relevant question, of course, is to what extent these results can be extended to nominally 
\textit{paraelectric} materials. As has recently been reported, even in a theoretically paraelectric material, an asymmetric distribution of defects can result in a small but measurable macroscopic piezoelectricity \cite{ref:Biancoli2015}. The reported effective piezoelectric coefficients for paraelectric perovskites is in the order of 0.05 C/m$^2$.
Compared to this average value, the same piezoelectric gradient of $10^{-2}$ C/m$^2$ across 1 mm required to yield $\mu_e =$ 1 $\mu$ C/m  represents to 20\% variation of the effective piezoelectric constant across the 1 mm thick beam. Though this gradient is large, it is not unrealistic.

The present analysis shows that piezoelectricity can indeed imitate flexoelectricity (bending-induced polarization)
on the condition that the piezoelectric coefficient be inhomogeneously and asymmetrically distributed across the sample. 
If this condition is met, however, asymmetric piezoelectricity - as might be found in bimorphs, but also as provided by surface piezoelectricity \cite{ref:Tagantsev2012,ref:Stengel2014,ref:Narvaez2015} - becomes 
functionally indistinguishable from intrinsic flexoelectricity. 
This perfect mimicry complicates the task of interpreting experimental results in flexoelectricity, but perhaps it also represents a practical opportunity; 
just like flexoelectricity was initially conceived as a way for replicating the device functionality of piezoelectrics \cite{ref:Zhu2006,ref:Fu2007,ref:Chu2009}, 
asymmetric piezoelectricity may be used to imitate the interesting novel functionalities \cite{ref:Lu2012,STARKOV201665,Cordero2017,Yangeaan3256} provided by flexoelectricity.

This research was funded by an ERC Starting grant (ERC 308023). 
All research in ICN2 is supported by the Severo Ochoa Excellence Programme (SEV-2013-0295) 
and funded by the CERCA Programme / Generalitat de Catalunya. F.V. thanks UCR, MICITT and CONICIT for support during his PhD. 

\bibliographystyle{apsrev4-1}
\bibliography{references}

\end{document}